\documentclass[runningheads]{llncs}
\usepackage{graphicx}
\usepackage{amsfonts}
\usepackage{mathabx}
\usepackage{multirow,diagbox}

\begin{document}
\title{Topic Modeling on Podcast Short-Text Metadata}
\titlerunning{Topic Modeling on Podcast Short-Text Metadata}
%\author{Francisco B. Valero\orcidID{0000-0002-2416-5163} \and Marion Baranes\orcidID{0000-0003-2385-727X} \and Elena V. Epure\orcidID{0000-0002-6930-9482}}
\author{Francisco B. Valero \and Marion Baranes \and Elena V. Epure}
\authorrunning{F. B. Valero et al.}
\institute{Deezer Research, 22-26, rue de Calais, 75009 Paris, France \\ \email{research@deezer.com}}
\maketitle              % typeset the header of the contribution
\begin{abstract}
Podcasts have emerged as a massively consumed online content, notably due to wider accessibility of production means and scaled distribution through large streaming platforms. 
Categorization systems and information access technologies typically use topics as the primary way to organize or navigate podcast collections.
However, annotating podcasts with topics is still quite problematic because the assigned editorial genres are broad, heterogeneous or misleading, or because of data challenges (e.g. short metadata text, noisy transcripts).
Here, we assess the feasibility to discover relevant topics from podcast metadata, titles and descriptions, using topic modeling techniques for short text.
We also propose a new strategy to leverage named entities (NEs), often present in podcast metadata, in a Non-negative Matrix Factorization (NMF) topic modeling framework.
Our experiments on two existing datasets from Spotify and iTunes and Deezer, a new dataset from an online service providing a catalog of podcasts, show that our proposed document representation, NEiCE, leads to improved topic coherence over the baselines.
We release the code for experimental reproducibility of the results\footnote{\url{https://github.com/deezer/podcast-topic-modeling}}.

\end{abstract}
\section{Introduction}
\label{sec:intro}
Podcasts is an audio content listened to on-demand for educational, entertainment, or informational purposes.
Known as the "spoken" version of blog posts, they have evolved towards a wide variety of formats (e.g. monologues, multi-party conversations, narratives) spanning a wide range of categories (e.g. business, true crime). 
Podcasts have been massively popularised in the recent years due to increased use of streaming platforms and availability of underlying technology for information access, recording and publishing  \cite{Jones2021current,Junta2008asimilar,Longqi2019more}.
As of summer 2021, the number of available podcasts in the world exceeds $2M$ and over $48M$ episodes \cite{winn2021podcaststats}.
Likewise, the podcast listening audience has grown massively: $50\%$ of the American population has listened to at least a podcast in 2019 and over $32\%$ have listened to podcasts monthly (compared to $14\%$ in 2014) \cite{edison2019}. 

Given the sharp growing interest in podcasts, researchers and industry players have searched for more suitable ways to enable listeners to effectively navigate these overwhelming collections \cite{besser2010podcast,clifton2020100000,Junta2008asimilar,Longqi2019more}.
Topics are central to any of the adopted podcast information access technology such as automatic categorization, search engines or recommender systems. 
However, annotating podcasts with topics is still quite problematic.
First, although podcast metadata entails topic-related genres, manually assigned by creators or providers, in reality these are often noisy and unreliable \cite{Jones2021current,sharpe2020review}: genres could be too heterogeneous or broad (e.g. Kids \& Family includes both sleep meditation and parenting advice);
and podcast providers may misleadingly label their shows with unrelated genres for broadening exposure.
Second, using topic modeling has its limitations inherited from the input text associated with podcasts:
1) metadata, such as title or description, is typically a short text of varying quality;
2) the automatically transcribed speech is noisy having a high word-error-rate especially for NEs and requires the engagement of more resources \cite{besser2010podcast,clifton2020100000,Jones2021current}.

In the current work, we revisit the feasibility of discovering relevant topics from podcast metadata, titles and descriptions, usually documented by podcast providers, creators, or editors.
While previous work \cite{besser2010podcast} found podcast metadata less promising for topic-informed search compared to when using podcast transcripts, we hypothesize that it could still be a very useful data source for topic modeling when exploited with appropriate technology.
If proven feasible, topic modeling on podcast metadata can be a more economic alternative than automatically extracting and exploiting transcripts of a rapidly growing podcast corpus.
Additionally, the discovery of topics directly from metadata offers many opportunities for improved podcast retrieval. 
Identifying podcast categories at different granularity levels could help editors evolve manually created podcast taxonomies and automatically annotate podcasts with these categories.
The discovered topics could also support the consolidation of podcast knowledge graphs \cite{benton2020trajectory,Jones2021current}, recently exploited in recommendation, by adding new edges that capture topic-informed podcast similarity based on metadata.

First, we take advantage of advancements in topic modeling, and benchmark multiple algorithms designed for short text on three podcast datasets.
Two of these datasets are public: one from Spotify \cite{clifton2020100000} and one from iTunes \cite{itunes}.
We have built a third dataset using Deezer\footnote{\url{https://www.deezer.com/us/}}, an online service providing a large podcast catalog.
This dataset is the largest with both titles and descriptions available at the podcast level.
Second, we propose a strategy to leverage NEs, frequently present in podcast titles and descriptions, in a NMF topic modeling framework.
As we can see in the following example: \textit{Shields Up! Podcast: Join Chris and Nev as they talk about their favourite Star Trek episodes covering everything from TOS to Lower Decks}, the metadata contains multiple NEs regarding the name of the speakers (\textit{Join Chris} and
\textit{Nev}), but also the podcast topic (\textit{Star Trek}, \textit{TOS} and \textit{Lower Decks}).
By injecting cues from NEs in topic modeling, we improve over state-of-the-art (SOTA) methods using plain word embeddings, and show that the data sparsity (very low co-occurrences of semantically related terms) due to short text can be further alleviated.

To sum up the contributions of this work are: a)~the most extensive study to date of topic modeling on podcast metadata, covering popular SOTA algorithms for short text and datasets from major podcast streaming platforms; b)~NEiCE, a new NE-informed document representation for topic modeling with NMF, as an extension of CluWords~\cite{viegas2019cluwords}---our approach improves topic coherence over baselines in most evaluated cases; c)~a new podcast dataset entailing English-language titles and descriptions from Deezer, an online service providing a podcast catalog, that is the largest in terms of the number of podcasts / shows.

\section{Related Work}
\label{sec:relatedwork}
Topic modeling on short text faces the challenge of severe data sparsity due to the nature of this type of input \cite{chen2019experimental}. 
Short text, as it consists of only few
words, can be ambiguous and noisy and, in general, has limited context.
This means that pairs of words that are topic-related do not or rarely co-occur in the same contexts, leading to conventional topic modeling techniques such as LDA \cite{blei2003we} to perform poorly.
Various topic modeling techniques have been designed to address this issue.
Models can be classified in four groups: pseudo-documents-based \cite{quan2015overwhelming,zuo2016topic}, probabilistic \cite{li2016topic}, neural \cite{miao2016neural,wu2020short}, and NMF-based \cite{shi2018short,viegas2019cluwords}. 
We further review each group and some representative models.

The principle of pseudo-documents is to aggregate connected short texts in longer documents, which are further used as input to conventional topic modeling \cite{li2016topic}.
Initial aggregation methods leveraged metadata such as hashtags in tweets \cite{Mehrotra2013Improving}.
However, this proved limiting for other types of short texts (e.g. search queries) and led to self-aggregation methods, able to aggregate using topic cues based on the corpus only \cite{quan2015overwhelming,zuo2016topic}.
An issue identified with this type of methods is overfitting \cite{zuo2016topic}.
Also, they appear overall less competitive than the other groups of topic modeling techniques for short text \cite{shi2018short,viegas2019cluwords}, discussed further.

The second group entailing probabilistic models is the most related to conventional topic modeling (LDA) that represents documents and topics as multinomial distributions over topics, respectively words. 
The adaptation of these models to short text is to assume that each document is sampled only from a single topic, thus restricting document-topic distribution to a mixture of unigrams \cite{nigam2000text,yin2014dirichlet,zhao2011comparing}. 
GPU-DMM \cite{li2016topic}, an effective and fast model in this group, is based on Dirichlet Multinomial Mixture (DMM) model and uses a Generalized Pólya Urna (GPU) as a sampling process to promote topic-related words.
The word association is estimated by exploiting pre-trained word embedding \cite{mikolov2013distributed}.
This allows to alleviate data sparsity as it extends the context to words that are semantically related but they do not necessarily co-occur in the same text.

The third group has become popular in the last years with the rise of deep learning. 
Neural topic modeling is based on Variational Auto-Encoders (VAE) \cite{bianchi2020pre,miao2016neural,srivastava2017autoencoding,wu2020short}.
Typically, an encoder such as a MultiLayer Perceptron (MLP) compresses the Bag-of-Words (BoW) document representation into a continuous vector.
Then, a decoder reconstructs the document by generating words independently \cite{bianchi2020pre,miao2016neural}.
Negative sampling and Quantization Topic Model (NQTM) \cite{wu2020short}, the latest topic modeling technique on short texts brings two contributions which yielded the current SOTA results.
The first is a new quantification method applied to the encoder's output whose goal is to generate peakier distributions for decoding.
The second is to replace the standard decoder with a negative sampling algorithm that proves better at discovering non-repetitive topics.

The NMF-based group learns topics by decomposing the term-document (BoW) matrix representation of the corpus into two low-rank matrices, one corresponding to document representations over topics and the other to topic representations over words \cite{kuang2015nonnegative}. 
Given the limited contextual information, the Semantics-assisted Non-negative Matrix Factorization (SeaNMF) model \cite{shi2018short} adjusts NMF to short texts by integrating into it word-context semantic correlations learnt from the skip-gram view of the input corpus.
In contrast to SeaNMF which focuses on the learning part, CluWords \cite{viegas2019cluwords} enhances the corpus representation before being factorized with standard NMF.
The matrix is obtained with a proposed custom TF-IDF strategy that exploits pre-trained word embeddings. 

The existing works include in their benchmark, datasets consisting of question or news titles, web snippets, review comments, or tweets \cite{he-etal-2018-interaction,li2016topic,shi2018short,viegas2019cluwords,wu2020short}.
Podcast metadata compared to these datasets exhibits a much higher frequency of NEs, which we exploit with the goal to further address data sparsity.
To our knowledge, we are the first to assess existing models on podcast metadata and to explicitly consider NE-related cues in short-text topic modeling.

%GPUDMM (2016) beats all the DMM methods and SATM (Quan 2015) which is one of the aggregating methods (which is slow on large corpora according to

%SeaNMF (2018) compared to LDA, NMF, PTM which is another aggregating methods (2016), GPUDMM (beats all of them) by large margins)

%CluWords (2019) compares with many probabilistic models including GPU-DMM and SeaNMF. It beats or is on par with SeaNMF in particular for very noisy text such as tweets

%NQTM (2020) compares with many probabilistic models including GPU-DMM and SeaNMF. Also with other neural models such as ProdLDA

\section{Methods}
\label{sec:methods}
The topic modeling algorithms we benchmark are GPU-DMM \cite{li2016topic}, NQTM \cite{wu2020short}, SeaNMF \cite{shi2018short} and CluWords \cite{viegas2019cluwords}.  
By noticing the high frequency of NEs in podcast titles and descriptions, we also include in the benchmark another standard NMF-based model for which we design a new NE-informed document representation as input.
The underlying hypothesis is that NEs convey the main topic information.
Thus, we propose to promote vocabulary words related to these NEs by associating them with pseudo-term frequencies as presented in Section \ref{sec:neinformed}.
For this, but also to capture word-to-word topic relatedness shown beneficial against data sparsity, we use pre-trained word and NE embeddings \cite{yamada2020wikipedia2vec}. 

Finally, the rationale behind choosing to explore NE promotion in a NMF framework is twofold. 
Compared to probabilistic models, NMF-based ones have yielded better results on short text \cite{chen2019experimental,shi2018short,viegas2019cluwords}.
Then, the integration of background NE and word information in NMF topic modeling is more straightforward than in deep neural networks.
Current autoencoders \cite{srivastava2017autoencoding,wu2020short} are designed to exploit only the corpus, which we find insufficient by itself to exhibit NE-word relations, especially if these corpora are small or each NE mention is infrequent.

\subsection{Notations and Preliminaries}
Table \ref{tab:notations} summarizes the notations used in the rest of the section.
As outlined above, we obtain topics by factorising the short-text corpus representation.
Formally, given the corpus $\mathcal{D}$, the vocabulary $\mathcal{V}$ consisting of unique words in $\mathcal{D}$, $A$ the matrix corresponding to BoW representations of each document in $\mathcal{D}$, and the target number of topics $K$, $A$ can be approximated by the product of two low-rank matrices $A \approx HW$.
Each row $W_{j,:}$ represents one of the K topics expressed in terms of words from $\mathcal{V}$ and each row $H_{i,:}$ represents an input document in terms of the learnt K topics.

\begin{table}
\caption{Notations used to present the topic modeling technique.} \label{tab:notations}
    \centering
    \begin{tabular}{|l|l|} \hline
       \textbf{Name}  & \textbf{Description} \\ \hline
       $K, k$  & Number of topics, the identifier of a single topic \\
       $\mathcal{D}, d$ & Short-text documents found in the corpus, a single document\\
       $\mathcal{V}, t, t^\prime$, $v_t$ & Vocabulary set, individual terms, the embeddings of term $t$\\
      $\mathcal{E}$, $e$, $v_e$ & Set of linked NEs, a NE term, the embedding of a NE $e$ \\
       $A \in \mathbb{N}^{|\mathcal{D}| \times |\mathcal{V}|}$ & Term-document matrix with BoW corpus representation \\
       $C \in \mathbb{R}^{|\mathcal{V}| \times |\mathcal{V}|}$ & Word similarity matrix computed with pre-trained embeddings\\
       $W \in \mathbb{R}^{K \times |\mathcal{V}|}$ & Latent low-rank word-topic matrix \\
       $H \in \mathbb{R}^{|\mathcal{D}|\times K}$ & Latent low-rank document-topic matrix\\
       $A^{*} \in \mathbb{R}^{|\mathcal{D}| \times |\mathcal{V}|}$ & Word-document matrix for CluWords corpus representation\\
       $A^{NE} \in \mathbb{R}^{|\mathcal{D}| \times |\mathcal{V}|}$ & Word-document matrix for NE-informed corpus representation\\
       $\alpha^{word}, \alpha^{ent} \in [0,1)$ & Minimum cosine similarity between words, or words and NEs \\
       \hline
    \end{tabular}
\end{table}

While this is the basic frameworks for NMF-based topic modeling, in practice there are more effective corpus representations than the simple BoW matrix ($A$), proven to lead to better topics.
CluWords \cite{viegas2019cluwords} is such an example and is based on two components:
1) one that correlates each word, not only with those with which co-occurs in the corpus, but also with other semantically related words, identified with the help of external pre-trained embeddings;
2) another one that derives a novel document representation, inspired by TF-IDF, which is able to incorporate information from the first component regarding word-to-word relatedness.
In our work, we choose to extend CluWords document representations to explicitly prioritize NE cues.
We further present the original CluWords, followed by the introduced changes in the next sub-section (\ref{sec:neinformed}).

The first step of CluWords is to compute a matrix $C$ where each element $C_{t,t^\prime}$ is the cosine similarity (cos) of the embeddings corresponding to the pair of terms $t, t^\prime \in \mathcal{V}$.
$C$ is constrained to be non-negative as it is used to compute $A^*$, which is the input to NMF. Thus, a positive cutoff $\alpha^{word}$ is used to select only the most similar term pairs, and nullify the rest of the matrix:
\begin{equation}
\label{eq:eq-1}
    C_{t, t^\prime} = \left\{
	\begin{array}{ll}
	\cos(v_t, v_t^\prime) & \textnormal{ if } \cos(v_t, v_t^\prime) > \alpha^{word} \\
	0 & \textnormal{ otherwise} \\
	\end{array}
	\right.
\end{equation}

Then, the BoW representation is replaced by a TF-IDF-inspired one.
Standard TF-IDF uses the corpus statistics to decrease the weight of very frequent terms and give more weight to terms that appear only in some contexts, thus judged more discriminative, while also accounting for term popularity in a document.
Equation \ref{eq:eq-2} shows how the TF-IDF score is computed for a term $t$ and a document $d$, where tf$(t, d)=A_{d,t}$ is the number of times $t$ appears in $d$ and $n_t$ is the number of documents in $\mathcal{D}$ where $t$ appears: 

\begin{equation}\label{eq:eq-2}
    \textnormal{tf\_idf}(t,d) = \textnormal{tf}(t, d) \cdot \log \left( \frac{|\mathcal{D}|}{n_{t}} \right)
\end{equation}

CluWords replaces $t$ by $C_{t,:}$ in order to avoid obtaining a very sparse representation matrix due to the limited context of each word in short text.
Thus, it redefines the tf and idf (the log ratio) from Equation \ref{eq:eq-2} to be computed over vector-based term representations instead of individual frequencies.
The new $tf^*$ and $idf^*$ in Equation \ref{eq:eq-3} incorporate information about semantically similar words to the term $t$ of a given document $d$ in order to expand the term's context:

\begin{equation}
    \label{eq:eq-3}
    A^*_{d,t} = \textnormal{tf}^*(d,t) \cdot \textnormal{idf}^*(t) = (AC)_{d,t} \cdot \log \left( \frac{|\mathcal{D}|}{\sum_{d\in\mathcal{D}}{\mu(t,d)}} \right) 
    \end{equation}

\noindent $\mu(t,d)$ is the mean cosine similarity between the term $t$ and its semantically related terms $t^\prime$ in document $d$ denoted $\mathcal{V}^{d, t} = \{t^\prime \in d | C_{t, t^\prime} \neq 0 \}$, or $0$ when the ratio in the first branch of Equation \ref{eq:eq-4} is undefined ($t$ is not in $d$, thus $|\mathcal{V}^{d, t}|=0$):
    \begin{equation}
    \label{eq:eq-4}
        \mu(t,d) = \left\{
        \begin{array}{ll}
      \frac{1}{|\mathcal{V}^{d, t}|} \cdot \sum_{t^\prime \in \mathcal{V}^{d, t} } C_{t, t^\prime} & \textnormal{ if }|\mathcal{V}^{d, t}|>0\\
      0 & \textnormal{ otherwise}
        \end{array}
        \right.    \end{equation}

\noindent Let us note that in the limit  case where C is the identity matrix, i.e. each term is only similar to itself which can be obtained by taking $\alpha^{word} =
\max_{t \neq t'}C_{t,t'}$, Equation (\ref{eq:eq-3}) becomes equivalent to Equation (\ref{eq:eq-2}). 
   
\subsection{NE-informed Corpus Embedding (NEiCE)}
\label{sec:neinformed}
Our approach NEiCE consists of a preprocessing step followed by a computation step which creates a new corpus representation matrix $A^{NE}$ leveraging NEs.

\subsubsection{Preprocessing step.} We identify NE mentions in podcast titles and descriptions and link them to Wikipedia entities using the Radboud Entity Linker (REL) system \cite{vanHulst2020rel}.
The REL system is based on multiple modules in pipeline specific to different sub-tasks: 1) the detection of NE mentions using Flair \cite{akbik2018coling}, a SOTA Named Entity Recognition (NER) framework using contextualized word embeddings;
2) the disambiguation of the identified entity against a list of possible Wikipedia candidates and its linking to the final candidate.
In this final linking phase, REL \cite{vanHulst2020rel} uses Wikipedia2Vec embeddings \cite{yamada2020wikipedia2vec}.

The Wikipedia2Vec embeddings that we also leverage in our solution, compared to other embeddings targeting words only \cite{mikolov2013distributed,mikolov-etal-2018-advances}, are learnt jointly for words and NEs from Wikipedia text. 
Their learning entails the optimization of three skip-gram sub-models  \cite{yamada2020wikipedia2vec}: 1) a regular word skip-gram; 
2) an anchor context model---for each NE mention appearing as a hyperlink in text its surrounding words become context;
and 3) a link graph model---the entities connected to a NE in the Wikipedia graph become context.
From all the information REL returns given a specific input, we use: the Wikipedia page of the disambiguated NE and the confidence score that helps us to choose if we treat a span of text as a NE or favour instead to process its words separately.

Finally, when NEs are processed as separate words instead of being linked to Wikipedia entities, we apply an extra vocabulary cleaning step.
As we noticed that in podcast metadata mentions of actors, athletes, or celebrities were very common and we want to avoid the extraction of topics focused on names, we remove these concerned words using the package NameDataset\footnote{https://github.com/philipperemy/name-dataset}.

\subsubsection{Computation step} 
We derive a new corpus representation matrix $A^{NE}$ as explained next.
If NEs are identified in a document with high confidence, then we exploit this information as the main topic-related cues.
One strategy to achieve this from previous work on regular text \cite{krasnashchok-jouili-2018-improving} is to favour NEs among the top words to describe topics.
Specifically, during preprocessing NEs are treated as n-gram terms and included in the vocabulary.
Then, re-weighting approaches are applied to these terms before being served as input to a standard or variations of LDA.
The idea behind re-weighting is to associate a larger pseudo-frequency (tf) to NEs such that they are more likely to be picked as topic descriptors.

Contrary to the above-mentioned approach, our goal is to take into account NEs without including them in the vocabulary. 
While indeed humans will find NEs very expressive to convey topics, this only happens if they already know them.
For popular NEs which typically appear in news data exploited in \cite{krasnashchok-jouili-2018-improving}, this would not necessarily pose a problem.
However, the NEs from podcast metadata tend to be less common or very specific to certain domains, hence less informative for humans trying to associate a topic label.
For instance, "That Peter Crouch Podcast" requires knowing that Peter Crouch is a footballer before being able to relate this podcast to football or sport.

The approach we propose is to still use re-weighting to boost NEs importance, but, instead of directly targeting NEs, focus on their semantically-related words. 
Let $\mathcal{E}^{e}=\{t| \textnormal{cos}(v_e,v_t) \geq \alpha^{ent}, \forall t \in \mathcal{V} - \mathcal{E} \}$ be the set of non-NE words from $\mathcal{V}$ most similar to a NE $e$.
Similar to when we computed $C$, a threshold $\alpha^{ent}$ is applied to fix a minimum cosine similarity value between a pair of Wikipedia2Vec embeddings involving a NE ($e \in \mathcal{E}$) and a word ($t \in \mathcal{V}$). 
Then, we still compute $A^{NE}$ with Equation \ref{eq:eq-3}, but replace $tf^*$ with $tf^{NE}$ as follows: 

\begin{equation}
    \label{eq:po-ned}
   \textnormal{tf}^{NE}_{d,t} = \left\{
    \begin{array}{ll}
	(AC)_{d,t} + \max_{t^\prime \in \mathcal{V}^{d,t}} (AC)_{d,t^{\prime}}& \textnormal{, if } t \in \mathcal{E}^{e} \textnormal{, } e \textnormal{ in } d \textnormal{ and } | \mathcal{V}^{d,t}|>0   \\
	(AC)_{d,t} & \textnormal{  otherwise} \\
\end{array}
\right.
\end{equation}
\noindent 
We chose to apply the NE-related re-weighting to the tf factor because we wanted to use NE-related words as the main signal for topics and the direct frequencies allowed us to have more control on it, as also emphasized by \cite{krasnashchok-jouili-2018-improving}.
Second, there are two branches depending on whether $t$ is a term very similar to a NE $e$ present in $d$.
If that is the case, a pseudo-frequency is computed by taking into account the maximum in the CluWords tf matrix ($\textnormal{tf}^*$) for a document $d$.
This means that the words related to a NE $e$ become either as important as the term with the largest weight ($t^\prime$) or more important if the word $t$ already appeared in $d$.

\section{Datasets}
\label{sec:datasets}
We start with describing the existing podcast datasets from iTunes \cite{itunes} and Spotify \cite{clifton2020100000}.
Then, we introduce our newly collected dataset, Deezer, which is the largest one among the three as shown in Table \ref{tab:datasets}.
All these datasets contain podcast metadata, titles and descriptions, in English-language.
Metadata is documented by providers or creators in an RSS feed, used by podcast aggregators and streaming platforms to make podcasts available to listeners.  
Although metadata exists for both podcasts (shows) and episodes within shows, we currently focus on shows as their information seemed more reliable.
By manually analysing episode metadata in the podcast catalog to which we had access, we noticed they often lacked description or inherited show description.

The iTunes dataset \cite{itunes} consists of 10 155 podcasts, popular at the moment of creation.
The Spotify dataset \cite{clifton2020100000} has 105 360 episodes sampled uniformly at random from podcasts proposed by professional creators (about 10\%) and by amateur creators (about 90\%).
The metadata of each episode contains the title and description of the parent show which we extract to create the final dataset used in the experiments.
From these two datasets, we keep podcasts with unique titles and with the concatenations of title and description longer than $3$ terms.
Additionally, for Spotify we select only the podcasts associated with the language identifiers "en" and "en-US".

\begin{table}
\caption{Summary of podcast datasets: the number of podcasts, the vocabulary size, the total number of NE mentions, the total number of podcasts with NEs in metadata, the mean number of words per title, and the mean number of words per description.} 
    \centering
    \begin{tabular}{|c|c|c|c|c|c|c|} \hline
      Dataset & $|\mathcal{D}|$ & $|\mathcal{V}|$ & \#NE mentions & \#podc. with NE & \#w/title & \#w/descr. \\ \hline
       Spotify      & 17~456 & 7~336 & 20~885 & 9~198 & 3.5 & 38.2 \\
       iTunes       &  9~859 & 7~331  & 24~973 & 6~994 & 4.9 & 56.4 \\
       Deezer   & 29~539 & 14~322 & 67~083 & 19~969 & 4.0 & 62.6 \\ \hline
    \end{tabular}
    \label{tab:datasets}
\end{table}

Deezer differs from the others in that it is the largest.
It covers 18 genres (Culture \& Society, Business, Films \& Games, Music \& Audio Commentary, Comedian, Sports, Education, Spirituality \& Religion, Information \& Politics, Health \& Fitness, Art, Entertainment, Lifestyle \& Entertainment, Stories \& Fiction, Science, Child \& Family, True Crime, and History), with a minimum of 300 podcasts per genre.
Although these categories are related to topics, as we previously discussed in Section \ref{sec:intro}, they tend to be broad and not always reliable.
We could notice a significant overlapping (e.g. Entertainment with Lifestyle \& Entertainment, Stories \& Fiction with True Crime, or Sports with Health \& Fitness), but also how a single category gathers multiple topics. 

To create the dataset we randomly sampled from the accessed collection, public podcasts which had titles and descriptions, and the language identifier "en".
As the language provided in the metadata was not always reliable, we also used two automatic language detectors, fastText \cite{joulin-etal-2017-bag} and CLD3 \cite{cld3}. 
We filtered out podcasts which were not found to be in English by both detectors.
Additionally, we also removed podcasts from unpopular genres ($<$300 shows).
Finally, we applied the same preprocessing as for the other two datasets.

Table \ref{tab:datasets} presents additional statistics of the used datasets.
All datasets contain a large number of NEs and we can find NE mentions in 50\%-70\% of the podcasts per dataset.
We can also observe that the average number of words per title is quite similar for all datasets, while the descriptions in Spotify tend to be shorter.

\section{Experimental Setup}
\label{sec:experimentalsetup}
We describe next the evaluation metric, the detailed preprocessing and experimental setup, and the environment we used for running the models.

We evaluated topic quality by relying on the widely used topic coherence \cite{roder2015exploring}.
A set of facts are said to have high coherence if they could support each other.
In topic modeling, this translates into mapping terms on facts and measuring the extent to which these terms tend to co-occur in corpora.
While the spectrum of word co-occurrence metrics for topic coherence is quite large \cite{Newman2010Automatic}, the exhaustive search performed in \cite{roder2015exploring} shows that $C_V$ correlates best with human judgement of topic ranking.
Thus, we decided to report $C_V$ scores in our evaluation.
Given a topic $k$ defined by its $T$ top words $t_1,t_2,...,t_T$, $C_V$ is defined as:

\begin{equation}
    \label{eq:cv-main}
    C_V(k) = \frac{1}{T} \sum_{i=1}^{T} \textnormal{cos}(v_{NPMI}(t_i), v_{NPMI}(t_{1:T}))
\end{equation}

\noindent $v_{NPMI}(t_i)$ and $v_{NPMI}(t_{1:T})$ yield two vectors computed with the Normalized Pointwise Mutual Information (NPMI) metric as follows:

\begin{equation}
\label{eq:cv-i}
    v_{NPMI}(t_i) = \big( \textnormal{NPMI}(t_i, t_j)\big)_{j=1,..,T}
\end{equation}

\begin{equation}
\label{eq:cv-t}
    v_{NPMI}(t_{1:T}) = \Bigg ( \sum_{i=1}^{T} \textnormal{NPMI}(t_i, t_j) \Bigg )_{j=1,..,T}
\end{equation}

\begin{equation}
    \label{eq:npmi}
    \textnormal{NPMI}(t_i, t_j) =  \frac{\log \frac{p(t_i, t_j)}{p(t_i)p(t_j)}}{-\log(p(t_i, t_j))}
\end{equation}
where $p$ is the probability of a term occurrence or co-occurrence in an external corpus.
We use Palmetto \cite{roder2015exploring} to compute $C_V$ for each topic $k$ on Wikipedia as external corpus, and average over all $K$ topics to obtain an aggregated value.

In all the reported experiments, we fix the number of top words $T$ to $10$ and vary the number of topics $K$ between $20$, $50$, $100$ and $200$.
During preprocessing, we keep all the linked NEs whose REL confidence score is higher than $0.9$ even if they only appear once in the corpus. For normal words, same as in  \cite{wu2020short}, we filter out from vocabulary those that appear less than $5$ times.
We also remove stop words using NLTK \cite{bird2009natural}.
The same preprocessing is applied before each topic modeling baseline.
We evaluate GPU-DMM \cite{li2016topic}, NQTM \cite{wu2020short}, SeaNMF \cite{shi2018short} and CluWords \cite{viegas2019cluwords} with their default hyper-parameters.
We assess the original CluWords with both fastText and Wikipedia2Vec embeddings \cite{mikolov-etal-2018-advances}.

As discussed in Section \ref{sec:methods}, NEiCE requires two parameters $\alpha^{word}$ and $\alpha^{ent}$.
\cite{viegas2019cluwords} motivates the choice of $\alpha^{word}$ between $0.35$ and $0.4$ in CluWords as it allows to select top 2\% of most similar pairs of words.
Compared to this approach which assumes $\alpha^{word}$ mainly dependent on the pre-trained embeddings, we investigate if it varies per dataset.
Thus, we test $\alpha^{word}$ with multiple values ($0.2$, $0.3$, $0.4$, $0.5$), where larger the value is, fewer words are selected as being semantically-related to a given term.
We proceed similarly for $\alpha^{ent}$. 
We run the experiments on an Intel Xeon Gold 6134 CPU @ 3.20GHz with 32 cores and 128GB RAM.

\section{Results and Discussion}
\label{sec:resultsanddsicussion}
The topic coherence scores obtained by the different topic modeling techniques for short text are presented in Table \ref{tab:baselines}.
First, we could notice that NMF-based methods (SeaNMF and CluWords) obtain the best scores in most of the cases.
Second, when comparing individual techniques, the ranking depends on the case (number of topics and dataset), but few trends emerge.
SeaNMF yields best topic coherence for the lowest number of topics (20) on two datasets.
Aligned with the previous literature \cite{viegas2019cluwords,wu2020short}, the SOTA models, NQTM and CluWords, obtain very often the best or second best scores, with CluWords ranking first in most cases (7/12).
These observations support our choices to work in a NMF framework and devise NEiCE as a CluWord extension, but informed by NEs. 

\begin{table}
    \centering
    \caption{Topic coherence scores ($C_V$ in \%) obtained by baselines on the three podcast datasets for 20, 50, 100 or 200 topics. CluWords is used with fastText embeddings and the default $\alpha^{word}=0.4$. Best scores are in bold and second best scores are underlined.}
    \begin{tabular}{|c|cccc|cccc|cccc|}\hline
   \multirow{2}{*}{\diagbox{Model}{Dataset}}  & \multicolumn{4}{c|}{\textbf{Deezer}} & \multicolumn{4}{c|}{\textbf{Spotify}} & \multicolumn{4}{c|}{\textbf{iTunes}} \\ \cline{2-13}
     & 20 & 50 & 100 & 200 & 20 & 50 & 100 & 200 & 20 & 50 & 100 & 200 \\ \hline
    \textbf{GPU-DMM} & 39.0 & 38.3 & 37.6 & 40.1 & 39.5 & 39.4 & \underline{39.7} & \underline{40.1} 
    & 39.6 & 38.5 & \underline{42.0} & 41.1 \\
    \textbf{NQTM} & 38.5 & \underline{42.2} & \underline{42.9} & \underline{45.8} & \underline{42.9} & \underline{41.6} & 39.3 & \textbf{40.2} & \textbf{48.4} & \textbf{46.6} & 38.2 & \underline{42.8} \\
    \textbf{SeaNMF} & \textbf{47.7} & 40.5 & 37.3 & 39.0 & \textbf{45.5} & 36.4 & 36.6 & 35.7 & 42.2 & \underline{41.8} & 35.1 & 36.9 \\
    \textbf{CluWords$_{ft}$} & \underline{39.7} & \textbf{44.0} & \textbf{46.3} & \textbf{54.5} & 40.2 & \textbf{42.3} & \textbf{43.4} & 39.5 & \underline{42.7} & 40.1 & \textbf{48.6} & \textbf{47.9}\\ \hline
    \end{tabular}
    \label{tab:baselines}
\end{table}

Table \ref{tab:cluwordswiki} shows the results for CluWords with Wikipedia2Vec words embeddings for different values of $\alpha^{word}$.
As mentioned in Section \ref{sec:experimentalsetup}, previously \cite{viegas2019cluwords} this parameter was fixed depending on the source of embeddings to 0.4 for fastText and 0.35 for word2vec.
However, no parameter sensitivity analysis was conducted, which we do now per dataset.  
We can see that the choice of $\alpha^{word}$: 1) has a significant impact on the results which could vary up to almost $12$ percentage points for Spotify, $K=50$;
2) is dependent on the assessed case (dataset, $K$) which previously was not considered; 
and 3) some values appear to emerge as better choices per dataset (e.g. $0.4$ for iTunes or $0.5$ for Deezer).

\begin{table}
    \centering
    \caption{Topic coherence scores ($C_V$ in \%) obtained by CluWords for different $\alpha^{word}$ values (0.2, 0.3, 0.4, 0.5) with Wikipedia2Vec embeddings on the three podcast datasets for $K \in \{20, 50, 100, 200\}$ topics. Best scores are in bold.}
    \begin{tabular}{|c|cccc|cccc|cccc|}\hline
   \multirow{2}{*}{Dataset}  & \multicolumn{4}{c|}{\textbf{Deezer}} & \multicolumn{4}{c|}{\textbf{Spotify}} & \multicolumn{4}{c|}{\textbf{iTunes}} \\ \cline{2-13}
     & 20 & 50 & 100 & 200 & 20 & 50 & 100 & 200 & 20 & 50 & 100 & 200 \\ \hline
    \textbf{CluWords$_{wk}$}(0.2) &  41.3 & 42.8 & 42.0 & \textbf{45.9} & 43.2 & \textbf{49.0} & 41.9 & \textbf{43.0} & 46.6 & 46.8 & 36.6 & 40.9\\
    \textbf{CluWords$_{wk}$}(0.3) & 39.8 & 41.3 & 45.6 & 44.1 & 42.8 & 37.8 & 46.4 & 37.8 & 44.6 & 40.7 & 39.0 & 40.3\\
   \textbf{CluWords$_{wk}$}(0.4) & 40.2 & 48.7 & 42.5 & 44.4 & \textbf{48.4} & 39.3 & 41.8 & 39.9 & \textbf{52.9} & \textbf{48.5} & \textbf{49.6} & 40.0 \\
    \textbf{CluWords$_{wk}$}(0.5) & \textbf{43.0} & \textbf{49.1} & \textbf{47.7} & 41.6 & 47.3 & 37.2 & \textbf{49.9} & 42.7 & 45.3 & 40.4 & 41.1 & \textbf{44.9} \\ \hline
    \end{tabular}
    \label{tab:cluwordswiki}
\end{table}

Further, we present in Table \ref{tab:ours} the topic coherence scores obtained with our proposed document representation, NEiCE, and different values of $\alpha^{word}$ and $\alpha^{ent}$.
First, we could notice that the introduction of NE cues has a positive impact and NEiCE obtains larger coherence scores than the baselines in most cases (datasets and numbers of topics).
The average of NEiCE increase over the best baseline scores is of 15.7\% for our best choice of parameters $\alpha^{word}$ and $\alpha^{ent}$, with a maximum increase of 37.7\% on Deezer and $K=50$. 
Additionally, the underlined scores in Table \ref{tab:ours}, which represent scores larger than those obtained by the baselines, show that, no matter the choice of $\alpha^{word}$ and $\alpha^{ent}$, NEiCE still yields better topic coherence in a majority of cases (85.4\%).
The most challenging case remains Deezer and $K=200$ in which only $\alpha^{word}=0.5$ and $\alpha^{ent}=0.3$ lead to a larger score than the best baseline, although the increase is small so most likely not significant statistically.

\begin{table}
    \centering
    \caption{Topic coherence scores ($C_V$ , in \%) obtained by NEiCE, our document embedding strategy, for different values of ($\alpha^{word}$, $\alpha^{ent}$) using Wikipedia2Vec embeddings on the three podcast datasets. Best scores per dataset and number of topic are in bold. Scores larger than all baselines presented in Table \ref{tab:baselines} are underlined.}
    \begin{tabular}{|c|cccc|cccc|cccc|}\hline
   \multirow{2}{*}{Dataset}  & \multicolumn{4}{c|}{\textbf{Deezer}} & \multicolumn{4}{c|}{\textbf{Spotify}} & \multicolumn{4}{c|}{\textbf{iTunes}} \\ \cline{2-13}
     & 20 & 50 & 100 & 200 & 20 & 50 & 100 & 200 & 20 & 50 & 100 & 200 \\ \hline
    \textbf{NEiCE} $(0.2, 0.3)$ & \underline{50.2} & \underline{48.9} & \underline{51.4} & 48.4 & \underline{51.7} & \underline{49.0} & \underline{45.2} & \underline{46.5} & \underline{49.3} & 43.3 & \underline{49.5} & 47.0  \\
    \textbf{NEiCE} $(0.2, 0.4)$ & \underline{53.1} & \underline{49.2} & \underline{50.8} & 50.6 & \underline{48.7} & \underline{48.7} & \underline{43.5} & \underline{41.7} & 47.2 & \underline{49.5} & \underline{\textbf{50.7}} & \underline{\textbf{51.3}}  \\
    \textbf{NEiCE} $(0.3, 0.3)$ & \underline{48.5} & \underline{52.1} & \underline{51.5} & 49.8 &  \underline{52.2} & \underline{49.0} & \underline{47.5} & \underline{47.6} & \underline{50.3} & \underline{\textbf{52.5}} & \underline{49.0} & \underline{48.2}  \\
    \textbf{NEiCE} $(0.3, 0.4)$ & \underline{53.3} & \underline{50.9} & \underline{\textbf{55.3}} & 51.6 & \underline{50.1} & \underline{48.5} & \underline{51.1} & \underline{\textbf{49.8}} & \underline{52.5} & \underline{49.5} & \underline{49.2} & \underline{49.8}  \\ 
    \textbf{NEiCE} $(0.4, 0.3)$ & \underline{53.2} & \underline{51.5} & \underline{52.2} & 50.0 & \underline{53.2} & \underline{49.5} & \underline{\textbf{50.5}} & \underline{45.9} & \underline{\textbf{52.8}} & \underline{50.1} & \underline{50.6} & \underline{51.1} \\
    \textbf{NEiCE} $(0.4, 0.4)$ & \underline{\textbf{56.4}} & \underline{52.6} & \underline{48.1} & 49.0 & \underline{51.0} & \underline{48.2} & \underline{47.3} & \underline{47.8} & \underline{52.4} & \underline{51.9} & \underline{49.9} & 47.4  \\
    \textbf{NEiCE} $(0.5, 0.3)$ & \underline{52.5} & \underline{56.3} & \underline{50.8} & \underline{\textbf{55.4}} & \underline{51.3} & \underline{47.7} & \underline{45.6} & \underline{45.4} & \underline{50.6} & \underline{46.5} & \underline{46.7} & \underline{49.0}  \\
    \textbf{NEiCE} $(0.5, 0.4)$ & \underline{56.3} & \underline{\textbf{60.6}} & \underline{54.9} & 53.3 & \underline{\textbf{55.0}} & \underline{\textbf{49.9}} & 46.7 & \underline{45.0} & \underline{50.5} & \underline{52.0} & \underline{48.7} & 46.1  \\
    \hline
    \end{tabular}
    \label{tab:ours}
\end{table}

From Tables  \ref{tab:cluwordswiki} and \ref{tab:ours}, we can notice that the best $\alpha^{word}$ in CluWords is not necessarily the best in NEiCE.
For instance, on iTunes, $\alpha^{word}=0.4$ was the best choice in Table \ref{tab:cluwordswiki}, while in Table \ref{tab:ours} $\alpha^{word}=0.2$ appears a better choice.
Also, the best pair of values for these parameters seems to depend largely on the case (dataset and K).
Thus, a grid search on a hold-out set is advisable with NEiCE.

\begin{table}
    \centering
    \label{tab:examples}
    \caption{Topics obtained with NEiCE or NQTM on Deezer and $K=50$.}
    \begin{tabular}{|c|c|c|}\hline
    k & \textbf{NEiCE} & \textbf{NQTM}\\ \hline
 
    1 & mindfulness, yoga, meditation, & psychotherapist, beirut, displays, \\ 
    
    & psychotherapy, psychotherapist, & remixes, weddings, adversity, namaste,
    \\ 
    
   & hypnotherapy, psychoanalysis, hypnosis,  &  kimberly agenda introducing \\ 
   
   & therapist, psychology & \\
    \hline
    
    2 & fiction, nonfiction, novel, author, & avenues, werewolf, criminal, pure,  \\
    
    & book, novelist, horror, cyberpunk, & imaginative, strategies, demand, \\ 
   
    & anthology, fantasy & agree, oldies, hang \\ \hline

   3 & republican, senator, senate, libertarian, & hour, sudden, key, genres, keeps,  \\
    
    & election, candidate, nonpartisan, &  round, neighbor, conservatives, \\
    
    & conservative, caucus, liberal & realize, fulfillment \\
    \hline
     
\end{tabular}
\end{table}

We selected some examples of topics obtained with NEiCE and NQTM\footnote{CluWords has similar top words as NEiCE for topics 1\&2 and did not find topic 3.} for Deezer and $K=50$ in Table \ref{tab:examples}.
We selected these topics considering the 18 genres introduced in Section \ref{sec:datasets} and assumed them likely related to Health \& Fitness (1), Stories \& Fiction or True Crime (2), and Information \& Politics (3).
Although NQTM yields more diverse top words, their association with a topic is less straightforward compared to NEiCE.
However, topic 2 in NQTM is clearly about True Crime, while in NEiCE could be also about Stories \& Fiction.

Finally, a qualitative analysis of the topics obtained with NEiCE on Deezer also revealed that many topics were related to world regions which, although easy to interpret, may be noisy if too frequent.
These results may be related to the podcasts' topics, but a more likely explanation is that region-related NEs are overweighted. 
Thus, a detailed study of NE weighting in NEiCE is still needed.

\section{Conclusion}
\label{sec:conclusion}
We presented a detailed study of topic modeling on podcast metadata covering popular SOTA techniques for short text.
Moreover, we proposed NEiCE, a new NE-informed document representation exploited in a NMF framework, and we showed it was more effective in terms of topic coherence than the baselines in various evaluation scenarios including three datasets (one of which, the largest, being newly released).
Future work aims to extend the study at the episode level, assess the document representation in downstream tasks, gain more insights into NEiCE especially in relation to the pre-trained embeddings and the choices of $\alpha s$, and conduct expert studies with editors to further validate mined topics. 

\bibliographystyle{splncs04}
\bibliography{mybibliography}

\end{document}